\documentclass[onecolumn]{aastex631}

\usepackage{bm}      
\usepackage{xcolor}  
\usepackage{amsmath}
\usepackage{enumitem}
\usepackage{hyperref}

\begin{document}

\title{Large Magnetic Flux Rope Formation in an X2.1 Flare observed on September 6, 2011}

\author[0000-0003-1363-3096]{Arpita Roddanavar}
\affiliation{Center for Solar-Terrestrial Research, New Jersey Institute of Technology, University Heights, Newark, NJ 07102-1982, USA}

\author[0000-0001-5121-5122]{Satoshi Inoue}
\affiliation{Center for Solar-Terrestrial Research, New Jersey Institute of Technology, University Heights, Newark, NJ 07102-1982, USA}

\author[0000-0001-9046-6688]{Keiji Hayashi}
\affiliation{Center for Computational Heliophysics, New Jersey Institute of Technology, University Heights, Newark, NJ 07102-1982, USA}

\author[0000-0002-8179-3625]{Ju Jing}
\affiliation{Center for Solar-Terrestrial Research, New Jersey Institute of Technology, University Heights, Newark, NJ 07102-1982, USA}

\author[0000-0003-2427-6047]{Wenda Cao}
\affiliation{Center for Solar-Terrestrial Research, New Jersey Institute of Technology, University Heights, Newark, NJ 07102-1982, USA}
\affiliation{Big Bear Solar Observatory, New Jersey Institute of Technology, 40386 North Shore Lane, Big Bear City, CA 92314-9672, USA}

\author[0000-0002-5233-565X]{Haimin Wang}
\affiliation{Center for Solar-Terrestrial Research, New Jersey Institute of Technology, University Heights, Newark, NJ 07102-1982, USA}

\begin{abstract}

Solar active region 11283 produced an X2.1 flare associated with a solar eruption on September 6, 2011. Observations revealed a pre-flare sigmoidal structure and a circular flare ribbon surrounding the typical two-ribbon structure, along with remote brightenings located at a considerable distance from the main flare site. To interpret these observations in terms of the three-dimensional (3D) coronal magnetic field dynamics, we conducted data-constrained magnetohydrodynamic (MHD) simulations. Using a non-linear force-free field (NLFFF) as the initial condition, we reconstructed a realistic pre-flare magnetic environment, capturing a sheared sigmoid above the polarity inversion line (PIL) surmounted by a fan-spine structure. Our simulations revealed that reconnection between the sigmoidal field, the adjacent fan-dome field lines, and the neighboring large loops facilitated the transfer of magnetic twist and led to the formation of a large magnetic flux rope (MFR). This transfer and propagation of twist are clearly visible throughout the MFR. As reconnection progresses, the entire fan–spine structure expands along with the evolving MFR. A notable outcome of the simulation is that the footpoints of the newly formed MFR align closely with the observed circular flare ribbon and the remote brightening region. Our findings suggest that a large MFR formed during the X2.1 flare, providing a coherent explanation for the observed phenomena.

\end{abstract}


\section{Introduction} \label{sec:intro}

Solar flares are sudden and rapid releases of free magnetic energy stored in the solar atmosphere manifesting as bursts of radiation that span the electromagnetic spectrum (Shibata \& Magara \citeyear{shibata_solar_2011}). Often, these flares are accompanied by coronal mass ejections (CMEs), massive eruptions of plasma and magnetic fields from the Sun’s corona into the heliosphere, with energies reaching up to 10$^{32}$ erg (Chen \citeyear{chen_coronal_2011}). Together, they are classified as eruptive events, whereas flares without associated CMEs are termed confined events. Understanding the dynamics of these phenomena is critical not only for fundamental plasma physics but also for their role in driving space weather, which impacts Earth's technological systems and environment (see, e.g., Gosling \citeyear{gosling_solar_1993}; Webb \& Howard \citeyear{webb_coronal_2012}). This requires a detailed exploration of their origins, their initiation mechanisms, and the development of predictive models.

The standard flare (CSHKP) model, developed by Carmichael \citeyear{1964NASSP..50..451C}, Sturrock \citeyear{1966Natur.211..695S}, Hirayama \citeyear{hirayama_theoretical_1974}, and Kopp \& Pneuman \citeyear{kopp_magnetic_1976}, was among the first to describe the origins of solar eruptions. This model, based on magnetic reconnection, explains how the restructuring of magnetic fields in a current sheet behind an erupting filament converts magnetic energy into kinetic and thermal energy. The standard model effectively accounts for the formation of parallel flare ribbons via electron acceleration along reconnected field lines in a two-dimensional (2D) framework. However, actual eruptions occur in a 3D space. To address this, 3D extensions of the standard model were developed through MHD simulations (e.g., Aulanier et al. \citeyear{aulanier_standard_2012}, \citeyear{aulanier_standard_2013}; Janvier et al. \citeyear{janvier_standard_2013}, \citeyear{janvier_electric_2014}).

Recent advancements in solar observations have facilitated 3D MHD simulations derived from photospheric magnetic field data. Since photospheric observations provide only the bottom surface information, non-linear force-free field (NLFFF) extrapolation is employed to model 3D coronal magnetic fields (see review of Inoue et al. \citeyear{inoue_magnetohydrodynamics_2016} and references therein). The “force-free” assumption is justified by the low coronal plasma beta (the ratio of gas pressure and magnetic pressure; Gary \citeyear{2001SoPh..203...71G}), which excludes plasma gas pressure and gravity. These extrapolated fields serve as initial conditions for data-constrained MHD simulations, bridging observations and theoretical models.

\begin{figure*}
\centering
\includegraphics[scale=0.8]{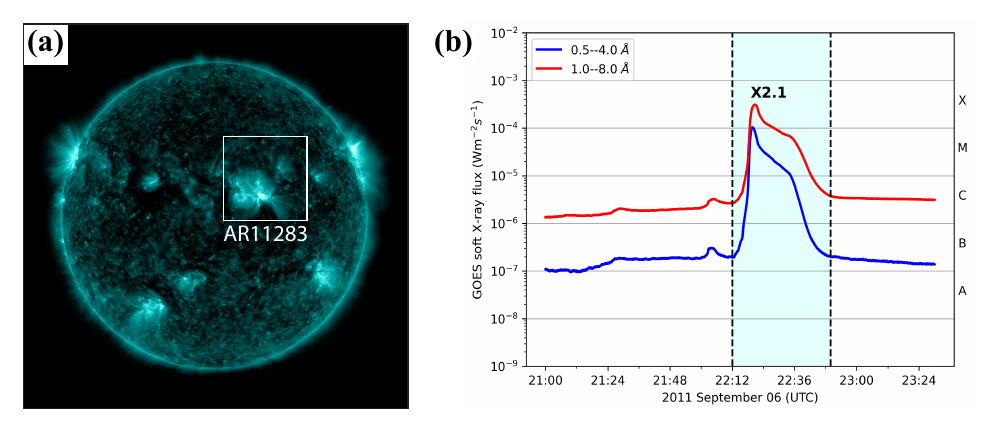} 
\caption{(a) Full-disk AIA 131 \AA\ image of the Sun during the X2.1 flare onset (SOL2011-09-06T22:18), with AR11283 enclosed within a white box.
(b) Time evolution of the GOES-15 satellite soft X-ray flux during the flare, recorded between 21:00 UT and 23:30 UT on September 6, 2011. The 0.5–4.0 \AA\ and 1.0–8.0 \AA\ passbands are plotted in blue and red, respectively. Vertical dashed black lines indicate the start and end times of the flare event.}
\label{fig:fig_1}
\end{figure*}

Eruptive events are often modeled by magnetic flux ropes (MFRs)- bundles of twisted magnetic field lines winding around a common axis. MFRs are considered key pre-eruptive magnetic structures; when destabilized, they form the core of a CME, driving its eruption from the solar surface. Observationally, CME progenitors manifest as filaments, prominences, hot channels, or sigmoids. Among these, sigmoids, S- or reverse S-shaped, highly sheared, and/or twisted magnetic structures in the lower corona (Rust \& Kumar \citeyear{rust_evidence_1996}) are regarded prominent precursors of solar eruptions (Canfield et al. \citeyear{canfield_sigmoidal_1999}, \citeyear{canfield_sigmoids_2000}; \citealt{2014IAUS..300..209G}) due to their exceptional ability to store and release free magnetic energy.

One prime example is active region AR11283, which exhibited a well-defined sigmoidal structure and was highly active during its disk passage, generating a series of recurrent M- and X-class flares, many of which were eruptive. This study focuses on the X2.1 flare that occurred on September 6, 2011, when AR11283 was positioned at N14W18. Owing to its intensity and location, this event has been extensively analyzed by numerous studies using various observational techniques (e.g., Wang et al. \citeyear{wang_relationship_2012}; Petrie \citeyear{petrie_abrupt_2012}; Dai et al. \citeyear{dai_production_2013}; Yang et al. \citeyear{yang_characteristics_2014}; Romano et al. \citeyear{romano_recurrent_2015}; Dissauer et al. \citeyear{dissauer_projection_2016}; Vanninathan et al. \citeyear{vanninathan_plasma_2018}) and modeling approaches (e.g., Feng et al. \citeyear{feng_magnetic_2013}; Liu et al. \citeyear{liu_three-dimensional_2014}; Janvier et al. \citeyear{janvier_evolution_2016}; Jiang et al. \citeyear{jiang_mhd_2013}, \citeyear{jiang_formation_2013}, \citeyear{jiang_nonlinear_2014}, \citeyear{jiang_data-driven_2016}, \citeyear{jiang_formation_2018}; Prasad et al. \citeyear{prasad_magnetohydrodynamic_2020}). 

In addition to its sigmoidal structure, AR11283 featured J-shaped parallel ribbons, followed by the appearance of revealing circular ribbons. First reported by Masson et al. (\citeyear{masson_nature_2009}), circular ribbons are, as their name suggests, closed and circular or elliptical in shape. Studies have shown that these circular ribbons are associated with a complex magnetic topology featuring a null point (NP, where the magnetic field strength vanishes), a fan-dome structure, inner and outer spine field lines, and remote brightenings linked to the outer spine (Wang \& Liu \citeyear{wang_circular_2012}). Unlike the 2D parallel ribbons in the standard model, where magnetic reconnection is thought to occur beneath the erupting MFRs, circular ribbon flares (CRFs) feature an authentic 3D magnetic configuration, with reconnection primarily occurring near the NP (Pontin et al. \citeyear{pontin_nature_2013}). Due to their unusual morphology and dynamic characteristics, CRFs do not conform to the ‘standard’ flare paradigm, making them a compelling subject for further investigation {\citep{2017A&A...604A..76M}}.

In this study, we employ NLFFF extrapolations and data-constrained 3D MHD simulations to investigate the X2.1 flare dynamics in AR11283. Our primary focus is to reproduce the eruption and correlate physical properties of the simulated 3D magnetic fields with key observed signatures, particularly, the sigmoid and the circular ribbons. This paper is structured as follows: Section \ref{sec:obs and methods} reviews the observational data and the methodology utilized for extrapolations and MHD simulations. Section \ref{sec:results} presents a detailed analysis of our simulation results and compares them with observations. Section \ref{sec:discussion} provides discussions. Finally, Section \ref{sec:summary} summarizes our conclusions and explores potential avenues for future research.



\section{Observations and Methodology} \label{sec:obs and methods}

\subsection{Observations} \label{subsec:obs}

Figure~\ref{fig:fig_1}(a) presents a full-disk image of the Sun captured in 131 \AA, with the active region (AR) of interest, AR11283 highlighted within a white box. The X2.1 flare commenced at 22:12 UT and reached its peak intensity around 22:21 UT, as indicated by the soft X-ray flux from the Geostationary Operational Environmental Satellite (GOES; Garcia \citeyear{1994SoPh..154..275G}), shown in Figure~\ref{fig:fig_1}(b). To analyze the early brightenings near the flare occurring region and the subsequent flux-rope evolution, we examine ultraviolet (UV) and extreme ultraviolet (EUV) data from the Atmospheric Imaging Assembly (AIA; Lemen et al. \citeyear{lemen_atmospheric_2012}) onboard the Solar Dynamics Observatory (SDO; Pesnell et al. \citeyear{pesnell_solar_2012}).  

To investigate the pre-flare magnetic field topology, we utilize the full-disk vector magnetogram series ``\texttt{hmi.B\_720s}" recorded at 20:36 UT on September 6, 2011, by the Heliospheric Magnetic Imager (HMI; \citealt{scherrer_helioseismic_2012}) aboard SDO. To obtain a larger field of view (FOV) than that provided by the standard Space-weather HMI Active Region Patches (SHARP; \citealt{bobra_helioseismic_2014}) data series, we generate our own Cylindrical Equal Area (CEA) projected maps using the same method as adopted in the SHARP pipeline. The resulting FOV spans $500'' \times 500''$, centered at heliographic coordinates ($260''$, $235''$), with a spatial resolution of $1''$ per pixel, selected to optimize computational efficiency while maintaining precision. The data are then 2 x 2 binned across the FOV, yielding a 250 × 250 grid that covers approximately 180 $\times$ 180 Mm$^2$ in physical space.


\subsection{Methodology} \label{subsec:methods}

We perform NLFFF extrapolation using the MHD relaxation method (\citealt{1994ApJ...422..899M}; \citealt{1994ASPC...68..225M}; \citealt{2011ApJ...727..101J}; \citealt{inoue_twist_2011}) and MHD simulations, both of which solve the zero-beta MHD equations (Inoue et al. \citeyear{inoue_magnetohydrodynamic_2014}) including:

\vspace{-5pt} 
\begin{equation}
\label{eq:rho}
\rho = |\bm{B}|,
\end{equation}

\vspace{-6pt} 

\begin{equation}
\label{eq:momentum1}
\frac{\partial \bm{v}}{\partial t} = \frac{1}{\rho} \bm{J} \times \bm{B} - \nu' \bm{v}
\end{equation}

\vspace{-6pt}

\text{or}

\vspace{-6pt}

\begin{equation}
\label{eq:momentum2}
\frac{\partial \bm{v}}{\partial t} = - (\bm{v} \cdot \nabla) \bm{v} + \frac{1}{\rho} \bm{J} \times \bm{B} + \nu \nabla^2 \bm{v},
\end{equation}

\vspace{-5pt}

\begin{equation}
\label{eq:induction}
\frac{\partial \bm{B}}{\partial t} = \nabla \times (\bm{v} \times \bm{B}) - \eta \nabla^2 \bm{B} - \nabla \phi,
\end{equation}

\vspace{-6pt}

\begin{equation}
\label{eq:current}
\bm{J} = \nabla \times \bm{B},
\end{equation}

\vspace{-6pt}

\begin{equation}
\label{eq:phi}
\frac{\partial \phi}{\partial t} + c_h^2 \nabla \cdot \bm{B} = - \frac{c_h^2}{c_p^2} \phi,
\end{equation}

Where $\rho$ is the mass density, $\bm{B}$ is the magnetic flux density, $\bm{v}$ is the velocity, $\bm{J}$ is the electric current density, and $\phi$ is a scalar potential used to minimize errors caused by divergence of the magnetic field ($\nabla \cdot \bm{B}$) in Equation~\eqref{eq:phi}, based on Dedner et al. (\citeyear{2002JCoPh.175..645D}). In Equation~\eqref{eq:rho}, the density evolution is linked to the magnetic field to facilitate the relaxation process by equalizing the Alfvén speed across space. The key physical variables, namely length, magnetic field, density, velocity, time and current density are normalized as follows: $L^*$ = 1.80 $\times$ 10$^8$ m, $B^*$ = 0.24 T, $\rho^*$ = $4.58 \times 10^{-8}$ kg/m$^3$, which is the density at the bottom surface of the simulation box, $V_A^* \equiv B^*/(\mu_0\rho^*)^{1/2}$ = 1.0 $\times$ 10$^6$ m/s, where $\mu_0$ is magnetic permeability in free space, $t_A^* \equiv L^*/V_A^*$ = 180 s, and $J^* \equiv (B^*/\mu_0L^*)$. $\nu'$ and $\nu$ represent the friction coefficient and viscosity, respectively. The resistivity $\eta$ was set differently for the NLFFF calculation and MHD simulations, with further details provided later. The advection and diffusion coefficients, $c_h^2$ and $c_p^2$ in Equation~\eqref{eq:phi}, were fixed at 0.04 and 0.01, respectively. 

In both NLFFF extrapolation and MHD simulations, the density is taken to be proportional to the magnetic field strength. With regards to the boundary conditions, the velocity field ($\bm{v}$) is set to zero on all boundaries, implying no photospheric driving. For the magnetic field, the treatment differs by boundary. At the top and lateral boundaries, the normal component is held fixed with time, while the tangential components evolve according to the induction equation (\ref{eq:induction}). At the bottom boundary, however, the magnetic field is treated differently in the NLFFF extrapolations and the MHD simulations, as described in the following Sections~\ref{subsubsec:nlfff} and \ref{subsubsec:mhd}. A Neumann-type boundary condition ($\partial_n \phi = 0$) is applied to the scalar potential $\phi$ on all boundaries, where $\partial_n$ denotes the derivative normal to the surface. The numerical scheme employed in all calculations approximates spatial derivatives using a second-order accurate central finite difference method in the interior, while one-sided forward and backward difference methods are applied near the boundaries. Time integration is carried out using the fourth-order Runge–Kutta–Gill method.

\subsubsection{NLFFF extrapolation} \label{subsubsec:nlfff}

To construct the NLFFF, referred to as Run 1, we first compute a potential field using the Green's function method \citep{1982SoPh...76..301S}, and use the 2D photospheric vector magnetic field as the bottom boundary condition for the extrapolation.

\begin{deluxetable*}{lcccccc} 
    \tablecaption{\centering An overview of each run for the MHD simulations, encompassing the equations for velocity evolution, resistivity, and the initial and boundary conditions on the transverse magnetic field components. “Relax” indicates that the transverse components evolve according to the induction equation (Equation(\ref{eq:induction})), while “Fix” denotes that they remain fixed in time.}\label{tab:runs}
    \tablehead{
        \colhead{Run} & \colhead{Type} & \colhead{Velocity Evolution} & \colhead{Resistivity} & \colhead{Initial Condition} & \colhead{Boundary Condition}
    }
    \startdata
    Run 2a & MHD            & Equation~\eqref{eq:momentum2} & Zero               & NLFFF           & Relax \\
    Run 2b & MHD            & Equation~\eqref{eq:momentum2} & Zero            & NLFFF           & Fix \\
    Run 2c & MHD             & Equation~\eqref{eq:momentum2} & Anomalous-Equation~\eqref{eq:ana res}             & NLFFF           & Fix \\
    Run 2d & MHD            & Equation~\eqref{eq:momentum2} & Zero               & $t=1.2$ in Run 2c & Relax
    \enddata
\end{deluxetable*}

At the bottom boundary, both the normal and tangential components of the magnetic field are held fixed. For velocity evolution, we solve Equation~\eqref{eq:momentum1} by replacing the nonlinear and viscous diffusion terms in Equation~\eqref{eq:momentum2} with a frictional term, following the MHD relaxation approach (e.g., \citealt{2011ApJ...727..101J};  \citealt{2020ApJS..247....6M}), where the friction coefficient $\nu'$ is set to 0.5. After establishing suitable initial and bottom boundary conditions, the equations are iteratively solved over time until the magnetic field reaches a near-equilibrium state. The resistivity in Equation~\eqref{eq:induction} is defined by:

\vspace{-5pt}

\begin{equation}
\label{eq:res}
\eta = \eta_0 + \eta_1 \frac{|\bm{J} \times \bm{B}||\bm{v}|^2}{|\bm{B}|^2},
\end{equation}

where $\eta_0$ = 5.0 $\times$ 10$^{-5}$ is the background resistivity (photospheric) and $\eta_1$ = 1.0 $\times$ 10$^{-3}$. The second term in this equation accelerates the force-free convergence of the magnetic field, particularly in weak-field regions.

The transverse magnetic field obeys the following equation to ensure a smooth transition between the inner region and the bottom surface, preventing sudden jumps:

\vspace{-5pt}

\begin{equation}
\label{eq:trans B}
\bm{B}_\text{BC} = \gamma \bm{B}_\text{obs} + (1 - \gamma) \bm{B}_\text{pot}.
\end{equation}

Here, the transverse component of the magnetic field ($\bm{B}_\text{BC}$) is a linear combination of the observed photospheric magnetic field ($\bm{B}_\text{obs}$) and the potential field ($\bm{B}_\text{pot}$). The coefficient $\gamma$ ranges between 0 and 1. In addition, we also introduce a force-free state indicator \textit{R}(=$\int \left| \bm{J} \times \bm{B} \right|^2 \, dV
$), which quantifies the Lorentz force integrated over the domain. When \textit{R} drops below a threshold value $R_{\text{min}}$, $\gamma$ is incremented as $\gamma = \gamma + \Delta\gamma$. Once $\gamma$ reaches 1, $\bm{B}_\text{BC}$ becomes fully aligned with $\bm{B}_\text{obs}$. In this study, we set $R_{\text{min}}$ = 1.0 $\times$ 10$^{-3}$ and $\Delta\gamma$ = 0.02.

Additionally, to prevent large velocity gradients between the inner region and the boundary surface, the velocity field is adjusted as follows:

\vspace{-5pt}

\begin{equation}
\label{eq:velocity}
\bm{v} \Rightarrow \frac{v_{\text{max}}}{v^*} \bm{v},
\end{equation}

If the Alfvén Mach number $v^*$ (given by $|\bm{v}|/|\bm{V}_A|$) exceeds $v_{\text{max}}$ (set to 0.04), the velocity is restricted to 4$\%$ of the Alfvén velocity.

\subsubsection{MHD simulations} \label{subsubsec:mhd}

Runs 2a-2c are MHD simulations initialized with the NLFFF. In Run 2a, the horizontal magnetic field components evolve according to the induction equation (\ref{eq:induction}) at the bottom boundary, while the normal component remains fixed over time. In contrast, for Runs 2b and 2c, all three magnetic field components are fixed over time at the bottom boundary. Unlike the NLFFF setup in Run 1, the velocity constraints (Equation (\ref{eq:velocity})) are removed for Runs 2a–2d.

For Runs 2a and 2b, the physical resistivity is set to zero (with numerical resistivity operating), while the non-dimensional viscosity is fixed at 1.0 $\times$ 10$^{-3}$. In Run 2c, the viscosity remains the same; however, anomalous resistivity is introduced (eg., Inoue et al. \citeyear{inoue_magnetohydrodynamic_2014}, \citeyear{inoue_magnetohydrodynamic_2015}), defined as follows:

\vspace{-5pt}

\begin{equation}
\label{eq:ana res}
\eta =
\begin{cases} 
\eta_0, & \text{if } J < J_c, \\
\eta_0 + \eta_2 \left(\frac{J - J_c}{J_c}\right)^2, & \text{if } J > J_c,
\end{cases}
\end{equation}

Where $\eta_0$ = 1.0 $\times$ 10$^{-5}$, $\eta_2$ = 1.0 $\times$ 10$^{-3}$ and $J_c$ is the critical current density above which anomalous resistivity becomes significant. In this study, $J_c$ = 40 was chosen through a process of trial and error.

Run 2d is an MHD simulation initialized from the magnetic field at $t=1.2$ obtained in Run 2c. Its boundary conditions and resistivity match those of Run 2a, except for the non-dimensional viscosity, which is reduced to 5.0 $\times$ 10$^{-4}$.

The type of simulation (NLFFF or MHD), velocity evolution equation, initial conditions, and boundary conditions for each run are summarized in Table~\ref{tab:runs}.

\begin{figure*}
\centering
\includegraphics[width=0.7\textwidth]{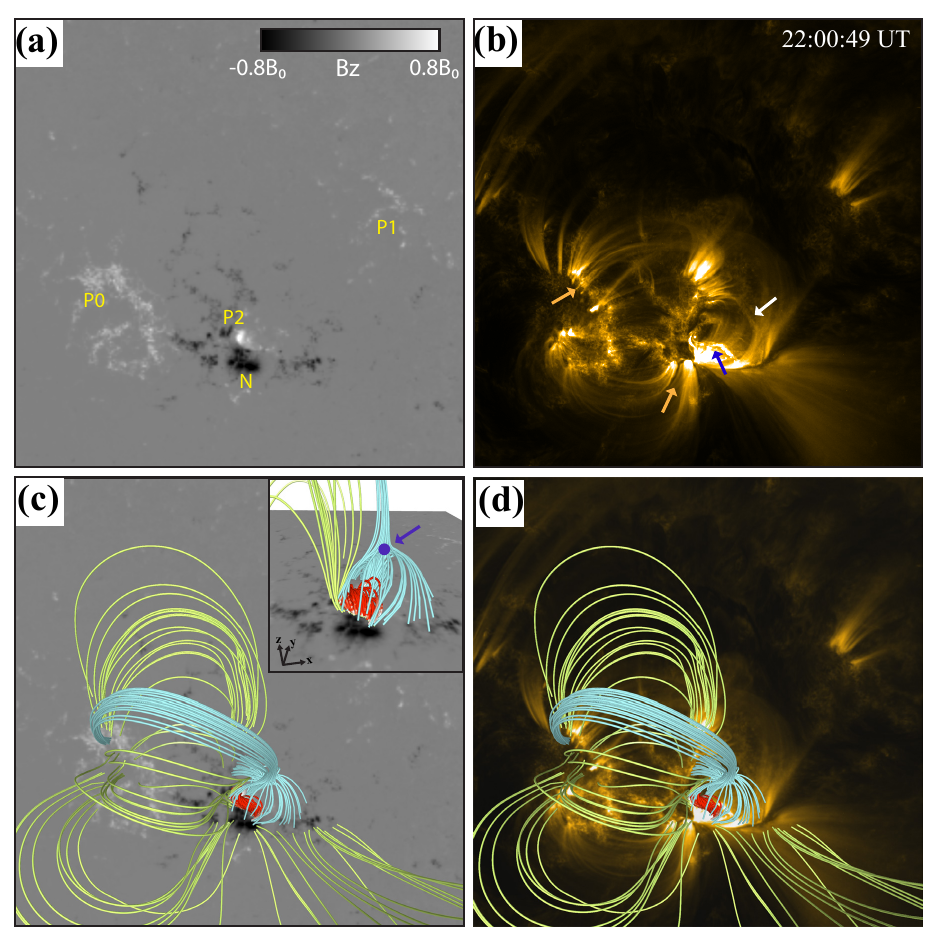}
\caption{(a) Photospheric radial magnetic field ($B_z$) of AR 11283 at 20:36 UT on September 6, 2011, used as the bottom boundary for all simulations. Positive and negative magnetic fields are shown in white and black, respectively, with $B_z$ normalized by $B_0$ (= 0.24 T). The four major magnetic polarities are labeled as P0, P1, P2, and N. The region spans approximately 180 × 180 Mm$^2$.  
(b) AIA 171 \AA\ snapshot of the same region near the flare onset time (22:00:49 UT). The blue arrow highlights an S-shaped brightening, the white arrow indicates circular loops above the S-shaped brightening, and the yellow arrows mark large neighboring coronal loops.  
(c) Top view of the extrapolated 3D coronal magnetic field lines illustrating the magnetic connectivity in the region. The red field lines represent the low-lying, sheared magnetic fields aligned along the polarity inversion line (PIL). The cyan field lines outline a dome-shaped structure connected remotely to P0, forming a fan-like topology. The yellow field lines trace the surrounding, large-scale coronal loops. The inset provides a zoomed-in view of the core magnetic structure, where the cyan fan-like field lines pass through a null point (NP) marked in purple.
(d) 3D magnetic field lines overlaid on the AIA 171 \AA\ image from panel (b).}
\label{fig:fig_2}
\end{figure*}

For all of the aforementioned runs, we use a 3D Cartesian computational box of 250 $\times$ 250 $\times$ 250 grid points ($\sim$180 × 180 × 180 Mm$^3$) with a resolution of 1 $\times$ 1 $\times$ 1 in non-dimensional units. This domain results from 2 $\times$ 2 binning of the original SHARP data (500 $\times$ 500 pixels). In this study, the simulation time $t$ is normalized by the characteristic Alfvén time $t_A^*$ = 3 minutes (see Section~\ref{subsec:methods}), so that $t$ = 1 corresponds to 3 minutes in physical time.  
\subsubsection{Quantitative analysis of the magnetic fields} \label{subsubsec:twist}

Magnetic Twist Number: This quantifies the twisting of the magnetic field lines and serves as a measure of how much the field deviates from a potential field state. This parameter is particularly useful for analyzing the evolution of the MFR and assessing its stability against kink instability (\citealt{1958PhFl....1..265K}; \citealt{1979SoPh...64..303H}; \citealt{torok_ideal_2004}). Therefore, to elucidate the evolution and dynamics of the MFR, we calculated the twist number according to the following equation (\citealt{berger_writhe_2006}):

\vspace{-5pt}

\begin{equation}
\label{eq:twist}
T_w = \frac{1}{4\pi} \int \frac{\bm{J} \cdot \bm{B}}{|\bm{B}|^2} \, dl,
\end{equation}

where $dl$ represents a line element. The twist was computed by tracing each field line from one footpoint to the other and plotted at the locations where the footpoints are anchored \citep{inoue_twist_2011}. Note that $T_w$ measures the number of turns of two infinitesimally close field lines (\citealt{berger_writhe_2006}), which is distinct from the number of turns of the field lines around the magnetic axis of the MFR \citep{2018SoPh..293...98T}. We traced the field lines starting one grid point above the photosphere, approximately 720 km, to minimize noise introduced by small-scale variations in the transverse magnetic field distribution.



\section{Results} \label{sec:results}

\subsection{Pre-Eruption 3D Magnetic Field Structure} \label{subsec:3.1}

\begin{figure*}
\centering
\includegraphics[width=0.95\textwidth]{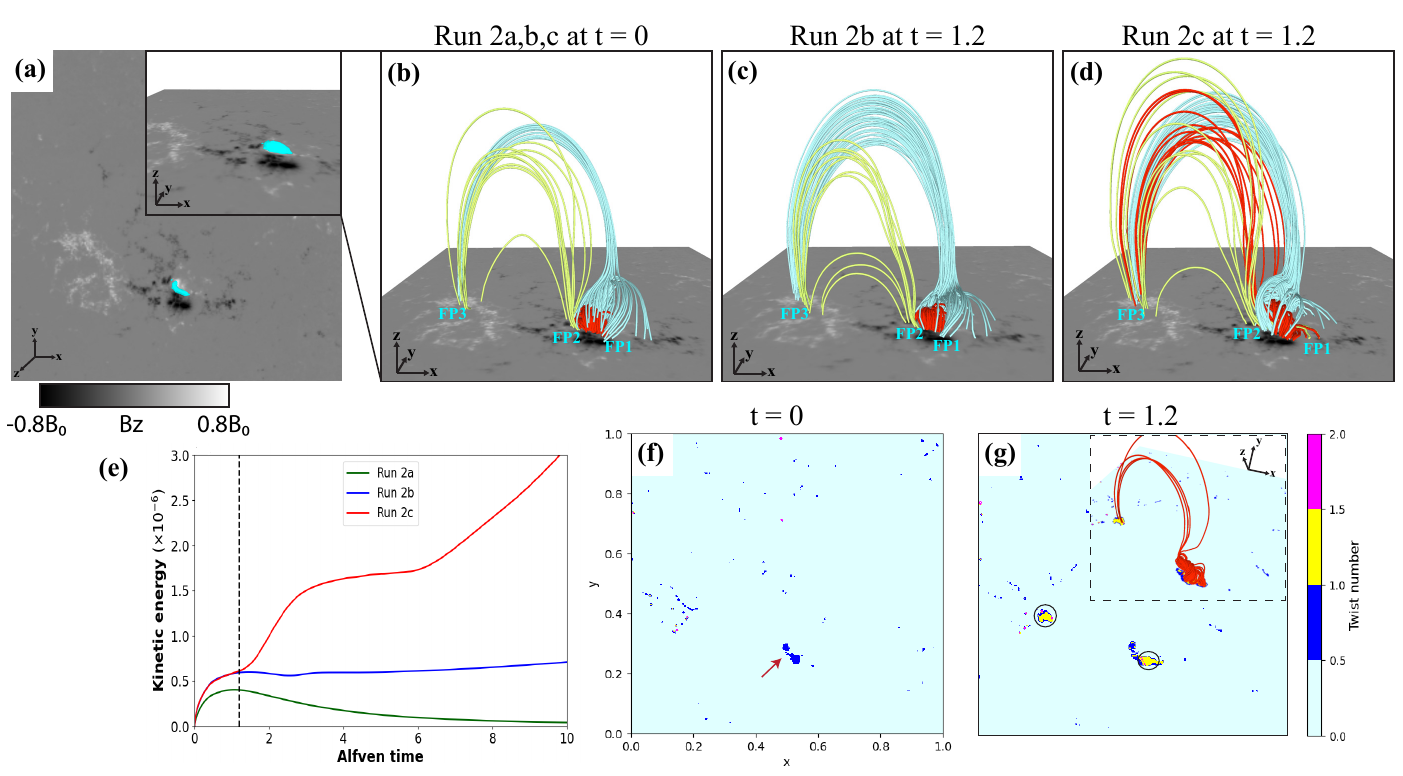}
\caption{(a) Iso-surface of current density ($|\bm{J}|) = 40$ in cyan, corresponding to the critical current in the anomalous resistivity, overlaid on the radial magnetic field ($B_z$). The inset provides a view of the same isosurface from a different field of view (FOV).
(b–d) Side views of the 3D magnetic field configuration: (b) NLFFF, (c) MHD simulation Run 2b at $t=1.2$, and (d) Run 2c at $t=1.2$, with footpoints (FPs) of the sigmoid marked in cyan.
(e) Temporal evolution of kinetic energy for Runs 2a (green), 2b (blue), and 2c (red). The vertical dashed black line marks $t=1.2$, where the magnetic fields from Run 2c serve as the initial condition for Run 2d.
(f, g) Magnetic twist maps corresponding to the field lines in (b) and (d), respectively, taken at approximately 720 km above the surface. In (f), the red arrow indicates a region with a twist number ranging from half a turn to one full turn. In (g), black circles mark regions where the twist exceeds one full turn. The inset panel presents field lines traced from these highlighted regions.}
\label{fig:fig_3}
\end{figure*}

Figure~\ref{fig:fig_2}(a) displays the photospheric vector magnetic field map derived from HMI at 20:36 UT, which serves as the bottom boundary condition for estimating the 3D coronal magnetic fields. The map highlights four major magnetic polarities: P0, P1, P2, and N, following the conventions established by \cite{prasad_magnetohydrodynamic_2020} and \cite{jiang_mhd_2013}. Figure~\ref{fig:fig_2}(b) shows a snapshot of the same region, captured in the AIA 171 \AA\ channel just before the flare at around 22:00 UT. Early signatures of S-shaped (sigmoid) hot loops, marked by a blue arrow are clearly visible along with bright loops encircling the sigmoid, indicated by a white arrow. Additionally, large circular bright loops appear to the east of the sigmoid, are marked by yellow arrows.

Figure~\ref{fig:fig_2}(c) illustrates the 3D magnetic field structure obtained from the NLFFF extrapolation, revealing distinct magnetic connectivity between the major polarities. Notably, strongly sheared, sigmoidal-shaped field lines (red) are reproduced between P2 and N, situated above the PIL. These field lines, which run parallel to the PIL and carry strong field-aligned currents, may play a crucial role in driving the eruption by facilitating the release of stored free magnetic energy \citep{jiang_mhd_2013}. The sigmoid lies beneath a 3D fan–spine structure (cyan), with the outer spine connecting to P0. The footpoints of the fan trace out a circular pattern on the bottom boundary and intersect one of the sigmoid’s footpoints. Additionally, large-scale magnetic loops (yellow) extend from P0 to N and reach into the upper corona, with some field lines extending close to the domain boundaries. The inset provides a zoomed-in view of the core magnetic topology near the PIL, showing the fan–spine field lines converging at a NP (indicated in purple), located at a height of approximately 13 Mm above the surface. These results are consistent with previous studies (e.g., \citealt{janvier_evolution_2016}). However,  \citet{prasad_magnetohydrodynamic_2020}, reported a slightly different topology, where the outer spine connects to P1 instead of P0. This discrepancy may stem from the use of different boundary conditions to model the coronal magnetic fields. While Prasad et al. (\citeyear{prasad_magnetohydrodynamic_2020}) implemented periodic boundary conditions, we use the Green's function method, {\it i.e.,} the magnetic field is assumed to be zero at infinity and outside the field of view (FoV). The differing boundary treatments could lead to variations in the resulting global magnetic field structure. Figure~\ref{fig:fig_2}(d) presents an overlay of the obtained magnetic field lines on the AIA 171 \AA\ snapshot. The global magnetic field configuration aligns closely with the observed brightenings in the AIA image (Figure~\ref{fig:fig_2}(b)).

\begin{figure*}
\includegraphics[scale=0.5]{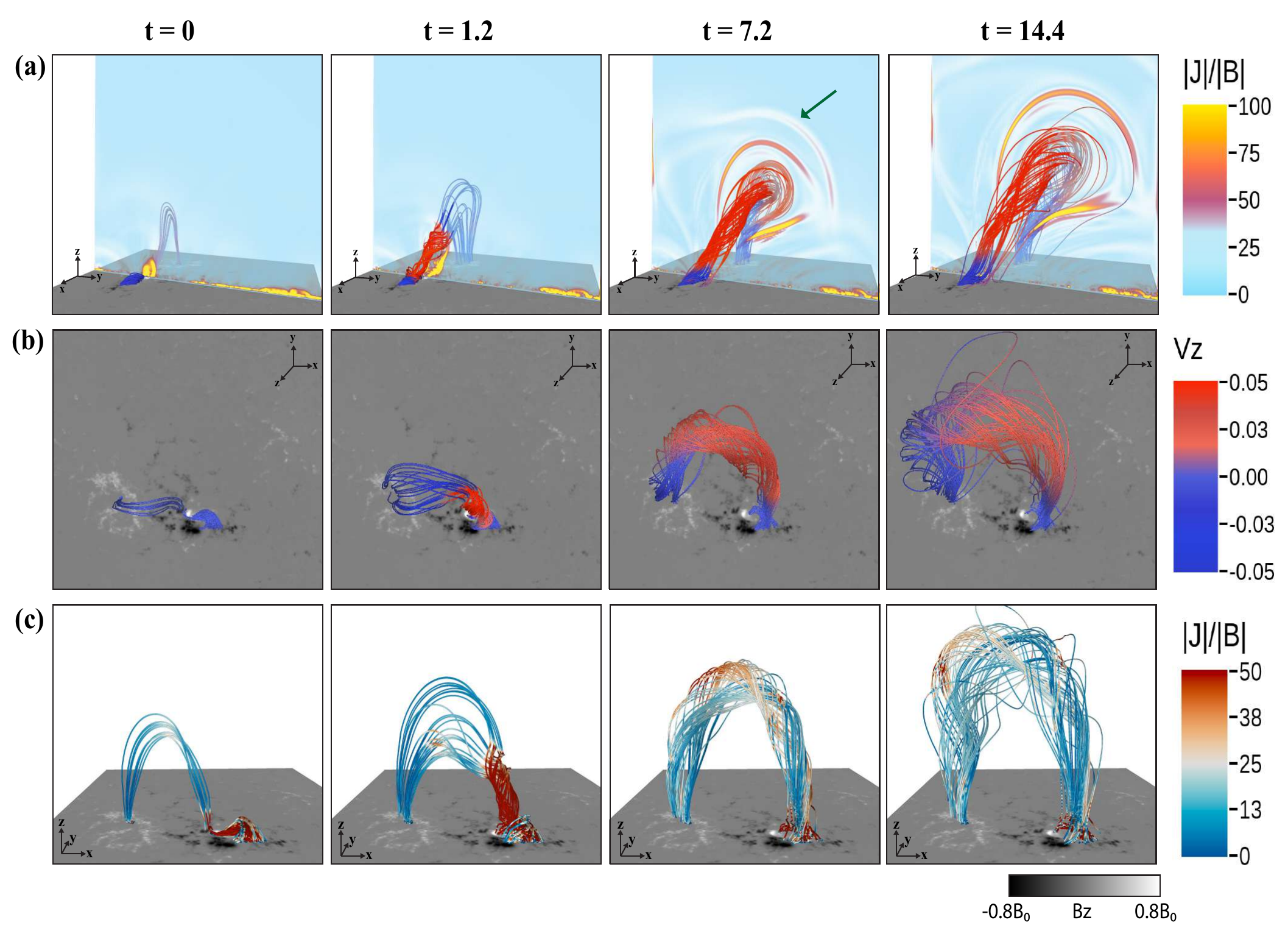} 
\caption{Formation and temporal evolution of the MFR during the data-constrained MHD simulation (Run 2d). (a) and (b) show side and top views of the magnetic field lines, respectively, colored by the vertical velocity component ($V_z$). The vertical slice in (a) displays the $|\bm{J}|/|\bm{B}|$ distribution. In panel (a), at \textit{t} = 7.2, the green arrow marks the presence of MHD waves ahead of the rising MFR. (c) Side view of the field lines, with $|\bm{J}|/|\bm{B}|$, illustrates the transfer of magnetic twist from the strongly twisted, low-lying structure to the large-loops during the MFR evolution. (An animation of this figure is available in the online journal. The composite animation shows the temporal evolution of the flux rope structure from $t = 0$ to $t = 14.4$, where $t = 1$ corresponds to 3 minutes in physical time. The realtime duration of the video is 6 s. It displays synchronized time evolution from panel (a) on the left, panel (b) in the center, and panel (c) on the right.)}
\label{fig:fig_4}
\end{figure*}

Thus, while the NLFFF approach is effective for constructing and analyzing the pre-eruption magnetic configuration, it also highlights the importance of understanding the complex magnetic connectivity within the active region. However, due to its static nature, it remains inadequate for capturing the highly dynamic behavior associated with eruptions. Therefore, an MHD simulation is necessary to accurately resolve the eruptive dynamics of the magnetic field.

\subsection{Magnetic Reconnection and MFR Formation} \label{subsec:3.2} 
We carry out MHD simulations to investigate the transition from a stable equilibrium to a dynamic state. In this context, we conduct data-constrained MHD simulations with the photospheric magnetic field as shown in Figure~\ref{fig:fig_3}(a) as the bottom boundary condition. 

We began with MHD simulations, referred to as Runs 2a and 2b, which differ in their boundary conditions (see Table~\ref{tab:runs}) but share the same initial condition provided by the NLFFF (Figure~\ref{fig:fig_3}(b)). The sigmoidal field lines (red) represent the non-potential magnetic component, whereas the fan-spine (cyan) and larger surrounding field lines (yellow) are nearly potential. Neither run resulted in an eruption, though their quantitative evolution differed (see Figure~\ref{fig:fig_3}(e)). By $t=1.2$, after the MHD simulation, the overall field line structure remained largely unchanged, as shown in Figure~\ref{fig:fig_3}(c).

Next, we introduced anomalous resistivity as defined in Equation~\ref{eq:ana res} in Run 2c. Figure~\ref{fig:fig_3}(a) shows an isosurface of strong current density ($|\bm{J}|$ = 40) in cyan, located above the PIL, corresponding to the region where the sigmoid formed. Starting with the NLFFF (Figure~\ref{fig:fig_3}(b)) as the initial condition, anomalous resistivity was implemented to enhance magnetic reconnection in high-current regions (\citealt{inoue_magnetohydrodynamic_2014, inoue_magnetohydrodynamic_2015}). Shortly after its introduction, reconnection occurred at the strong current region near FP2, where part of the sigmoid reconnected with the adjacent fan-dome field lines (cyan). This allowed the reconnected fields to expand beyond the fan-dome and subsequently interact and reconnect with neighboring large loops (yellow), leading to the formation of twisted field lines within the sigmoid (Figure~\ref{fig:fig_3}(d).)

Figure~\ref{fig:fig_3}(e) illustrates the temporal evolution of the kinetic energy profile for Run 2a (blue), Run 2b (green), and Run 2c (red). Initially, all three profiles exhibit a slight increase in kinetic energy until $t=1$ because the NLFFF is not fully in equilibrium. After $t=1$, the green line gradually declines, representing the relaxation of magnetic fields to a lower energy state due to the released boundary conditions. In contrast, the blue line follows a different trend, showing a small increase in kinetic energy after $t=3$ as a result of the constraints imposed by fixed boundary conditions. Meanwhile, the red line displays a significant rise in kinetic energy over time, indicating a transition from a near-equilibrium state to a dynamic one driven by reconnection. Consequently, at the onset of this dynamic phase (marked by the dashed black line) around $t=1.2$, we selected the magnetic field configuration for further analysis, with a detailed explanation provided later.

To quantitatively analyze the change in the magnetic field state between the NLFFF, which serves as the initial magnetic field of Run 2c, and the field at $t=1.2$ in Run 2c, we computed and compared the magnetic twist for each field line. Figure~\ref{fig:fig_3}(f) shows that at $t=0$, the magnetic twist, mapped onto the bottom surface, ranges from half a turn to one turn and is localized in the sheared field regions, as indicated by a red arrow. Since the twist number of these field lines is significantly below the critical threshold for kink instability ($T_n \geq 1.5 \, \text{--} \, 2.0$; Fan \& Gibson \citeyear{fan_emergence_2003}, Török et al. \citeyear{torok_ideal_2004}), the possibility of kink instability can be excluded. Note that their definition of twist differs from ours; however, the twist given in Equation (\ref{eq:twist}) serves as a useful proxy for assessing kink instability \citep{2016ApJ...818..148L}. However, strong-twist regions appear at \textit{t} = 1.2, exhibiting a twist exceeding one turn (yellow regions in Figure~\ref{fig:fig_3}(g)). This increase in twist results from enhanced reconnection within the sheared fields in high-current regions above the PIL as seen in Figure~\ref{fig:fig_3}(d). Additionally, reconnection between the sigmoid and the large loops allows twist to propagate along the loops, appearing near footpoint FP3. Field line tracing from these high-twist regions (marked by black circles in Figure~\ref{fig:fig_3}(g)) reveals a structure (inset) resembling the configuration in Figure~\ref{fig:fig_3}(d), confirming that these regions correspond to footpoints of post-reconnection field lines.

\begin{figure}
\centering
\includegraphics[width=0.45\textwidth]{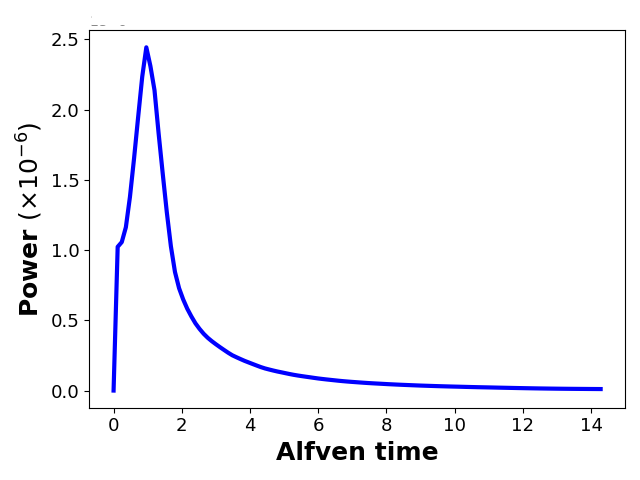}
\caption{Temporal evolution of the magnetic-to-kinetic energy conversion rate (${\bm v}\cdot({\bm J}\times {\bm B})$), computed as the volume integral ($\int \bm{v} \cdot (\bm{J} \times \bm{B}) \, dV
$), plotted as a function of normalized Alfvén time.}
\label{fig:fig_5}
\end{figure}

\subsection{Evolution and dynamics of the MFR} \label{subsec:3.3} 

Following the results of Run 2c, we infer that anomalous resistivity-driven magnetic reconnection plays a key role in generating strongly twisted fields that are no longer in equilibrium (\citealt{LiuN2025}). However, as noted earlier, the fixed boundary conditions and zero velocity at the bottom boundary make it inconsistent for the simulation to run for longer periods, as they violate the induction equation (Equation~\ref{eq:induction}). Therefore, we initiated Run 2d, with the twisted magnetic fields at $t=1.2$ obtained from Run 2c as the initial condition. In this run, the horizontal magnetic field components ($B_x$ and $B_y$) evolve according to the induction equation, while the normal component remains fixed in time. Note that this evolution is no longer consistent with the observations.

\begin{figure*}
\centering
\includegraphics[width=1.0\textwidth]{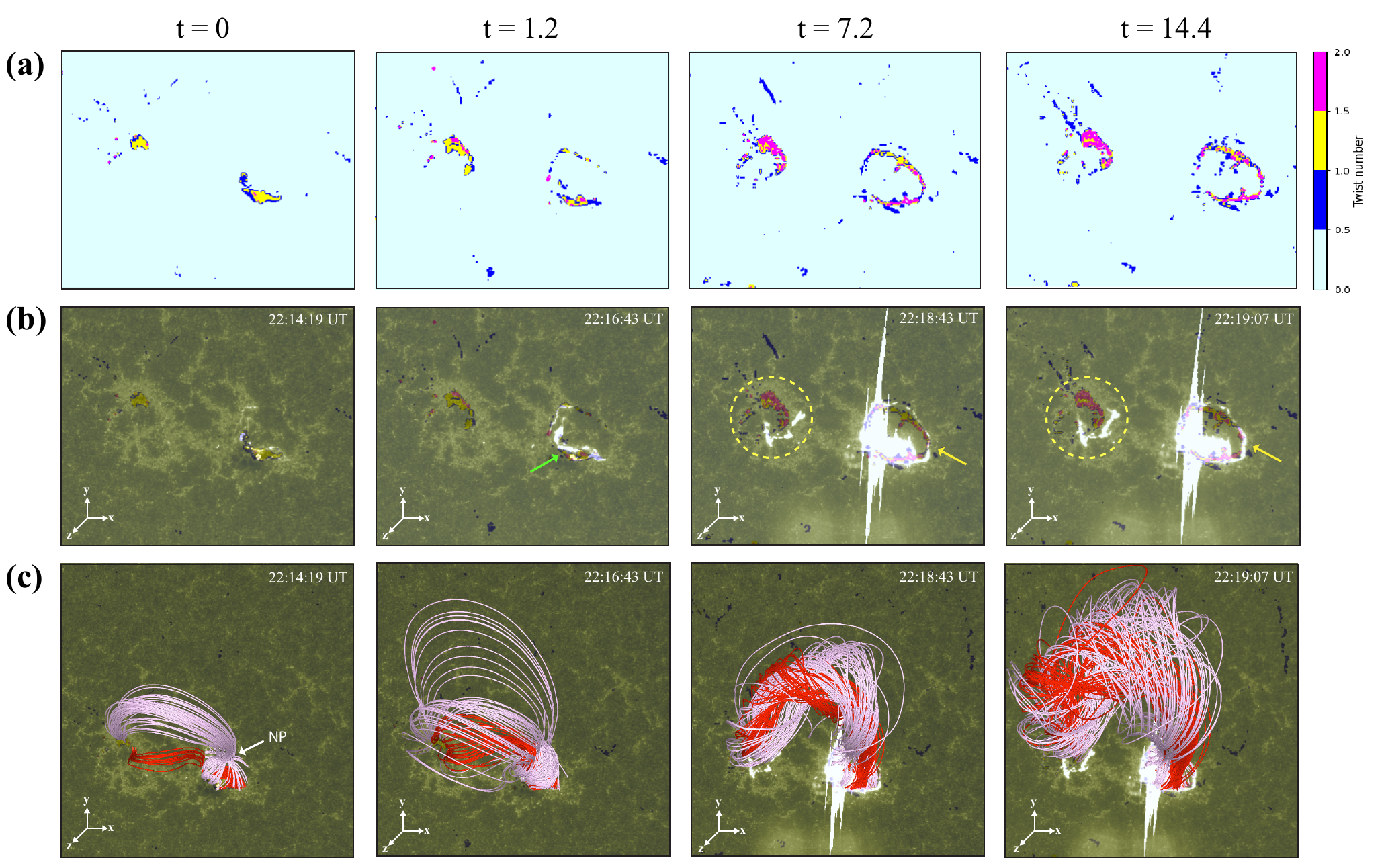} 
\caption{(a) Temporal evolution of magnetic twist during the MHD simulation, Run 2d, showing the transition from a localized twist distribution to a more circular pattern over time. The color bar represents the magnetic twist number, ranging from 0 to 2.0. 
(b) AIA 1600 \AA\ snapshots illustrating the progression of flare ribbons, from initially J-shaped parallel ribbons (indicated by a green arrow) to circular-shaped ribbons (indicated by yellow arrows). Dashed yellow circles mark remote brightenings.
(c) Temporal evolution of the large MFR, featuring a fan-spine magnetic topology (in pink) and highly twisted sigmoidal field lines having reconnected with the neighboring loops (in red) overlaid on the corresponding AIA 1600 \AA\ images from panel (b) and twist maps from panel (a). At \textit{t} = 0, NP (white arrow) marks the location of the null point.}
\label{fig:fig_6}
\end{figure*}

In Run 2d, we analyzed the temporal evolution of the magnetic field lines, focusing particularly on the evolution of the twisted fields within the sigmoid. Figures~\ref{fig:fig_4}(a) and \ref{fig:fig_4}(b) show the normal component of velocity ($V_z$) over the field lines, along with a vertical slice depicting the distribution of $|\bm{J}|/|\bm{B}|$ from the side and top perspectives, respectively. We observe that the twisted field lines, having already merged with the large loops through reconnection, propagate their twist along the loops, forming a large flux rope that rises and moves away from the surface. Since the newly formed MFR has an upward velocity concentrated at its top, we find that the magnetic energy accumulated in the sigmoidal twisted field lines is converted into kinetic energy within the large MFR. This interpretation is further supported by Figure~\ref{fig:fig_5}, which shows the temporal evolution of the magnetic-to-kinetic energy conversion rate (${\bm v}\cdot({\bm J}\times {\bm B})$), calculated over the entire 3D simulation domain. The profile demonstrates that this process predominantly occurs during the early reconnection phase. As the MFR rises, its evolution becomes inclined relative to the vertical direction, with an enhanced $|{\bf J}|/|{\bf B}|$ surrounding the MFR, as seen in Figure~\ref{fig:fig_4}(a). This inclination is similar to the eruption modeled by Kang et al. (\citeyear{kang_data-driven_2024}) for an M5.3 flare in the same active region earlier that day. Further analysis of the key factors driving such MFR rise is discussed later in the paper.

\subsection{Magnetic Twist Transfer into Large Loops} \label{subsec:3.4} 

The evolution and significance of magnetic twist in erupting flux ropes have been examined through both observations (eg., Wang et al. \citeyear{wang_witnessing_2015}; Pal et al. \citeyear{2021A&A...650A.176P}) and modeling efforts (Prasad et al. \citeyear{prasad_magnetohydrodynamic_2020}; Fan et al. \citeyear{fan_data-driven_2024}). Figure~\ref{fig:fig_4}(c) shows the 3D magnetic field lines with $|\bm{J}|/|\bm{B}|$, viewed from a different angle. Note that $|\bm{J}|/|\bm{B}|$ is used to detect strong magnetic twist in the weak magnetic field regions. This representation clearly illustrates that the twist of the sigmoid is eventually transferred to the system of quasi-potential loops via reconnection, in about 10 Alfv\'en times during Run 2d. To gain a deeper insight, we map the twist of the field lines onto the bottom surface, as shown in Figure~\ref{fig:fig_6}(a). The twist distribution evolves from a fairly concentrated region around the sigmoid at \textit{t} = 0 to a broader, dome-shaped region by \textit{t} = 14.4. Furthermore, high-twist regions appear to the east, near the footpoints of the large loops, providing further evidence for the propagation of twist along the MFR length.

\subsection{Comparative Analysis of Modeling Results and Observational Data} \label{subsec:3.5} 

In this section, we compare the results of our simulation with observable signatures during and after the eruption to validate the accuracy of our modeling.

\subsubsection{Flare ribbons, twist maps and MFR evolution} \label{subsubsec:3.5.1} 

Figure~\ref{fig:fig_6}(b) presents an overlay of the twist map with AIA 1600 \AA\ snapshots taken shortly before the flare. The AIA images capture the evolution of J-shaped parallel ribbons near the sigmoid (marked by a green arrow), which later extend northward, forming circular ribbons (indicated by yellow arrows). Additionally, remote brightenings are observed east of the sigmoid location and are marked by yellow circles. At \textit{t} = 1.2, twist distribution approximately traces the lower regions of the parallel two-ribbon structures. The twist evolution, which also expands out in an almost circular pattern as seen in the final panels of Figure~\ref{fig:fig_6}(a), exhibits excellent spatial resemblance to the structure and evolution of the circular flare ribbons. Moreover, portions of the remote flare ribbons, highlighted by dashed yellow circles, partially coincide with regions of strong twist.
Prasad et al. (\citeyear{prasad_magnetohydrodynamic_2020}) suggested the formation of circular flare ribbons and standard 2D flare ribbons through subsequent reconnections at the 3D NP and X-point, respectively. 
In the context of examining the spatial relationship between the twist distribution, flare ribbons, and the magnetic field structure, we trace the field lines corresponding to both the parallel and circular flare ribbons and overlay them with the twist map and AIA 1600 \AA\ images in Figure~\ref{fig:fig_6}(c). In addition to the highly twisted sigmoidal field lines and the neighboring large loops (red), the fan–spine field lines (pink) connected to the the NP (indicated by the white arrow) are also present. Initially, at \textit{t} = 0, the pink field lines exhibit no twist, resembling potential fields. Immediately after, these pink lines interact with the red ones close to the NP and the twist is transferred via reconnection. Note that the evolution of the red field lines results from reconnection between the sigmoidal fields and the outer fan-dome field lines, which enabled them to expand beyond the fan structure and subsequently reconnect with the large coronal loops. As a result, twist is transferred through the NP and across the entire fan–spine structure, allowing the spine and the NP to become integrated into an even larger flux rope, with footpoints that spatially align with the twist map, one anchored at the circular ribbons and the other located close to the remote flare ribbons on the eastern side. \citet{janvier_evolution_2016} reported a similar correlation between these remote flare brightening and 3D quasi-separatrix layers (QSLs), suggesting long-range field-line interactions, particularly associated with the NP configuration (Masson et al. \citeyear{masson_nature_2009}). Our simulation provides compelling evidence that the circular flare ribbon and remote brightening regions are anchored at the footpoints of the newly formed MFR.

\begin{figure}
\centering
\includegraphics[width=0.47\textwidth]{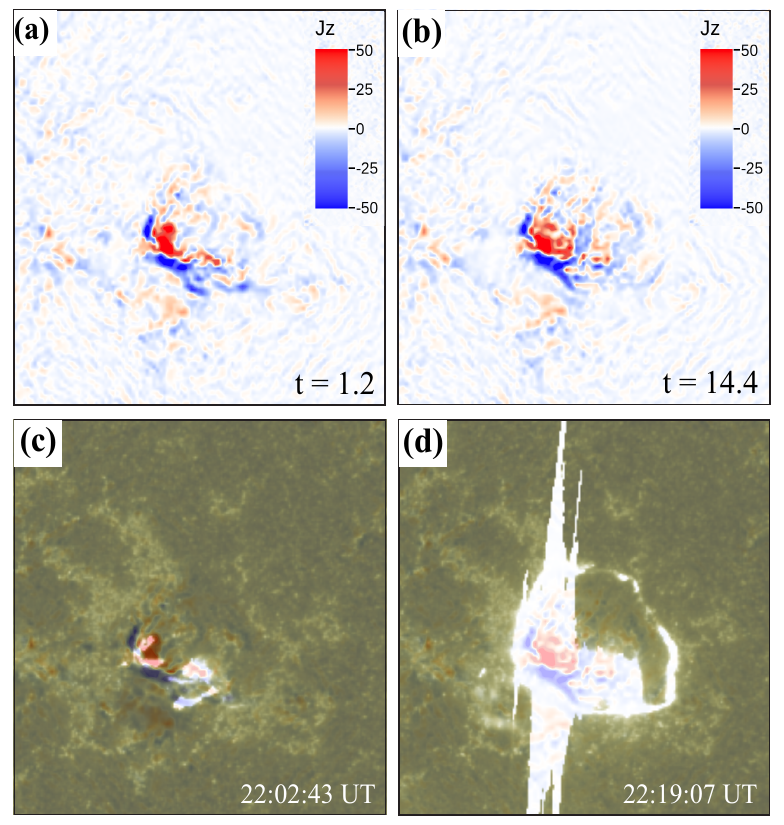}
\caption{Distribution of strong vertical current density ($J_z$) in the active region near the flare ribbon locations. Panels (a) and (b) show $J_z$ distribution at \textit{t} = 1.2 and \textit{t} = 14.4, respectively, during the MHD simulation (Run 2d). Panels (c) and (d) display AIA 1600 \AA\ snapshots at times corresponding to the appearance of parallel and circular flare ribbons, respectively. These snapshots are overlaid on the $J_z$ maps from panels (a) and (b), highlighting the spatial correlation between the current density and the flare ribbons.}
\label{fig:fig_7}
\end{figure}

\subsubsection{Vertical current density (Jz) map } \label{subsubsec:3.5.2} 

Janvier et al. (\citeyear{janvier_electric_2014}) pointed out the intensification of the vertical current density ($J_z$) during flares, which originates from the curl of the horizontal magnetic field, $(\nabla \times \bm{B})_z$, on the photosphere. They compared the distribution and evolution of the currents with those of the flare ribbons using observational data and a 3D simulation model (\citealt{aulanier_standard_2012}). Later, \citet{janvier_evolution_2016} analyzed the same active region as in our study and observed a direct alignment between vertical currents and the J-shaped 2D parallel ribbons, further corroborating their earlier findings. While previous studies have examined the topology and temporal evolution of vertical currents before and after flares (\citealt{mitra_circular_2023}) and their correlation with H$\alpha$ ribbons and emission regions  (\citealt{sharykin_fine_2014}), changes in $J_z$ associated with circular ribbons remain largely unexplored. We compute and present the vertical current density ($J_z$) distribution near the flare ribbon locations in Figure~\ref{fig:fig_7}. Figures~\ref{fig:fig_7}(a) and ~\ref{fig:fig_7}(b) illustrate the corresponding $J_z$ distributions at \textit{t} = 1.2 and \textit{t} = 14.4, respectively. We overlay AIA 1600 \AA\ images on the $J_z$ map around the time of the appearance of the parallel and circular flare ribbons in Figures~\ref{fig:fig_7}(c) and ~\ref{fig:fig_7}(d), respectively. The strong $J_z$ (in red) distribution aligns with the hooks of the J-shaped parallel ribbons, consistent with the findings of Janvier et al. (\citeyear{janvier_evolution_2016}). However, no significant enhancement of $J_z$ is observed near the circular ribbons, even in regions of strong twist, as seen in the last panel of Figure~\ref{fig:fig_6}(b). Figure \ref{fig:fig_6}(c) shows that the circular flare-ribbons correspond to one of the anchoring footpoints of the MFR. As the MFR rises, the horizontal magnetic field at the bottom surface weakens, leading to a decrease in $J_z$. This result indicates that the vertical current density is not always enhanced at flare ribbon locations.

\begin{figure}
\centering
\includegraphics[width=0.45\textwidth]{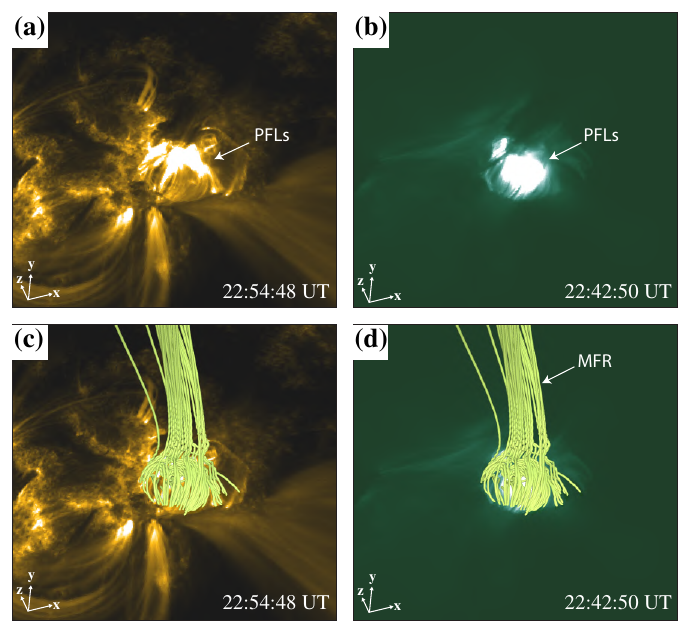} 
\caption{(a) and (b) AIA 171 \AA\ and 94 \AA\ images showing post-flare loop-like (PFLs) configurations marked by white arrows, respectively.
(c) and (d) Magnetic field lines overlaid on AIA 171 \AA\ and 94 \AA\ images, respectively, illustrating the presence of similar post-flare loop-like structures beneath the MFR, also marked by the white arrow.}
\label{fig:fig_8}
\end{figure}

\begin{figure*}
\centering
\includegraphics[width=0.9\textwidth]{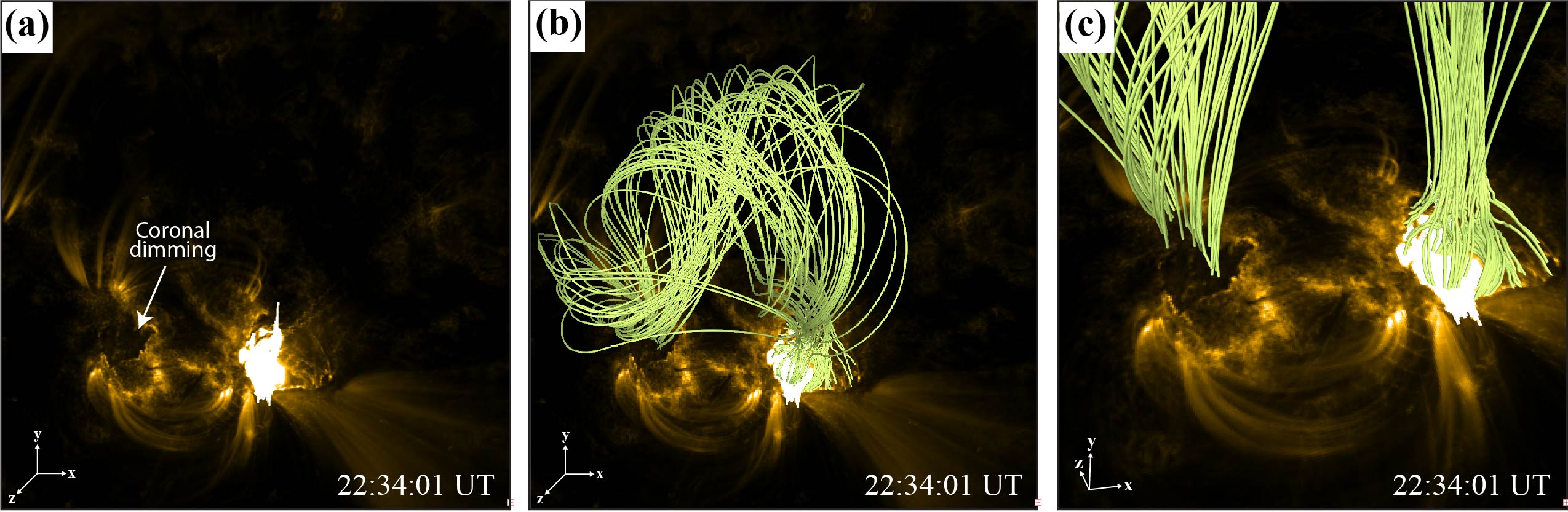} 
\caption{(a) AIA 171 \AA\ image showing a prominent coronal dimming region (white arrow) observed during the eruption. 
(b) MFR (yellow) overlaid on the image from panel (a). 
(c) Zoomed-in view of panel (b), showing the eastward footpoint of the MFR, which coincides with the dimming region.}
\label{fig:fig_9}
\end{figure*}

\subsubsection{Post-flare loops and coronal dimmings} \label{subsubsec:3.5.3}

We further compare the observed compact post-flare loops (PFLs) with simulation results. Figures~\ref{fig:fig_8}(a) and ~\ref{fig:fig_8}(b) highlight these compact PFLs observed in the AIA 171 \AA\ and 94 \AA\ channels, respectively, revealing bright loop tops likely caused by interactions between adjacent loops (Smartt et al. \citeyear{smartt_post-flare_1993}). In the final panels of Figures~\ref{fig:fig_4}(b) and ~\ref{fig:fig_6}(c), small loop-like structures are observed to form near the westward footpoints of the erupting MFR across the PIL. To further analyze this, we plot the corresponding field lines and overlay them with the AIA 171 \AA\ and 94 \AA\ images in Figures~\ref{fig:fig_8}(c) and ~\ref{fig:fig_8}(d) respectively. Our results indicate that the compact PFLs correspond to the footpoint of the erupting MFR, and the strong brightening was likely caused by reconnection occurring at the NP within the fan-spine topology.

\begin{figure*}
\centering
\includegraphics[width=0.9\textwidth]{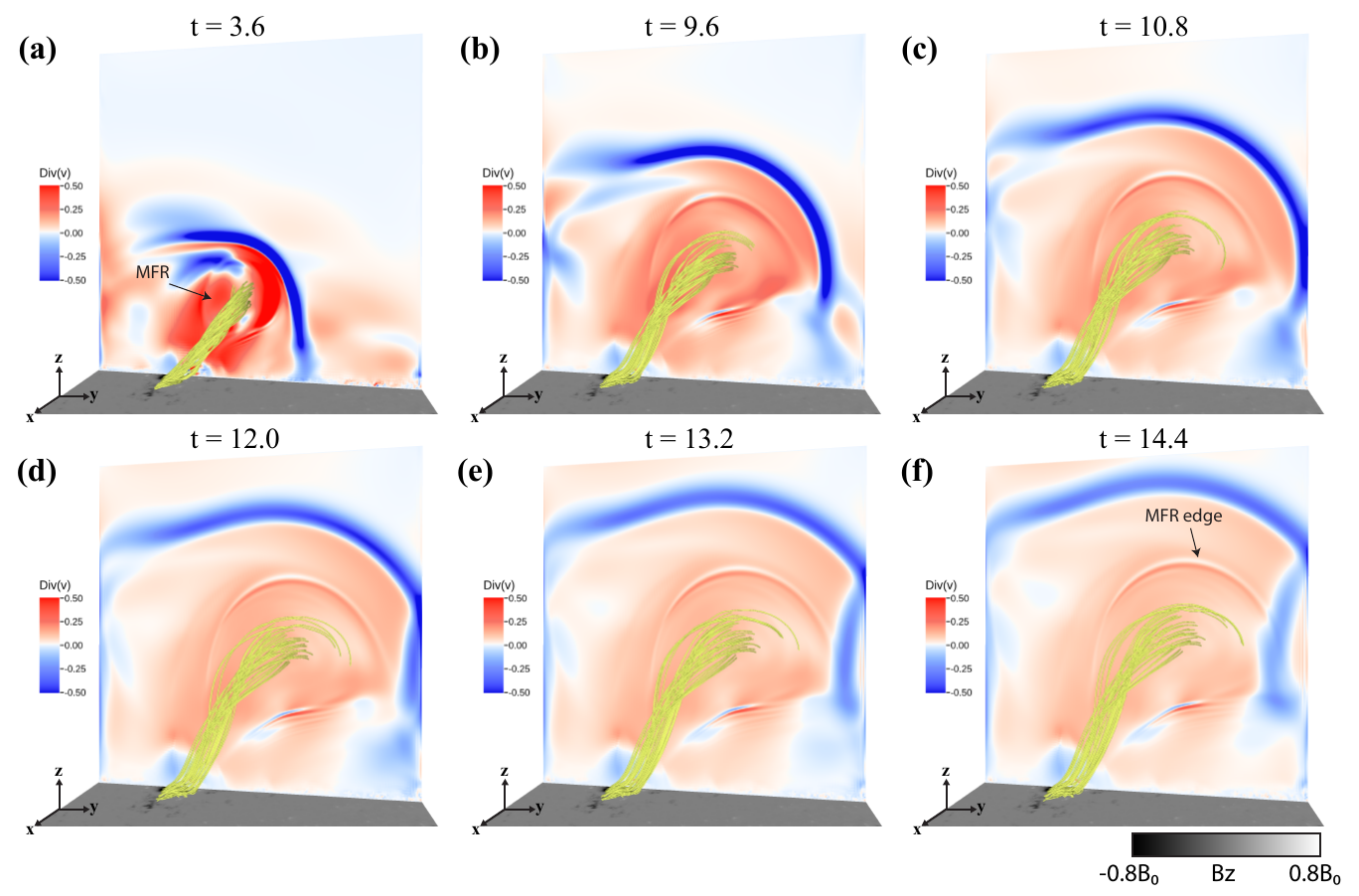} 
\caption{(a-f) Temporal evolution of the MFR (in yellow), with the divergence of velocity ($\nabla \cdot \bm{V}$) shown on the vertical cross-section in the y-z plane. Blue and red colors represent compressive and expanding waves, respectively. (f) The black arrow marks the MFR edge interacting with compressible waves reflected off the simulation box boundary. (An animation of this figure is available in the online journal, showing the continuous evolution of the MFR from $t = 0$ to $t = 14.4$, where $t = 1$ corresponds to 3 minutes in physical time. The realtime duration of the video is 6 s. The animation reveals how the interaction between compressible waves reflected from the simulation box boundary and the MFR edge influences the slow-rising phase of the MFR.)}
\label{fig:fig_10}
\end{figure*}

\begin{figure}
\centering
\includegraphics[width=0.47\textwidth]{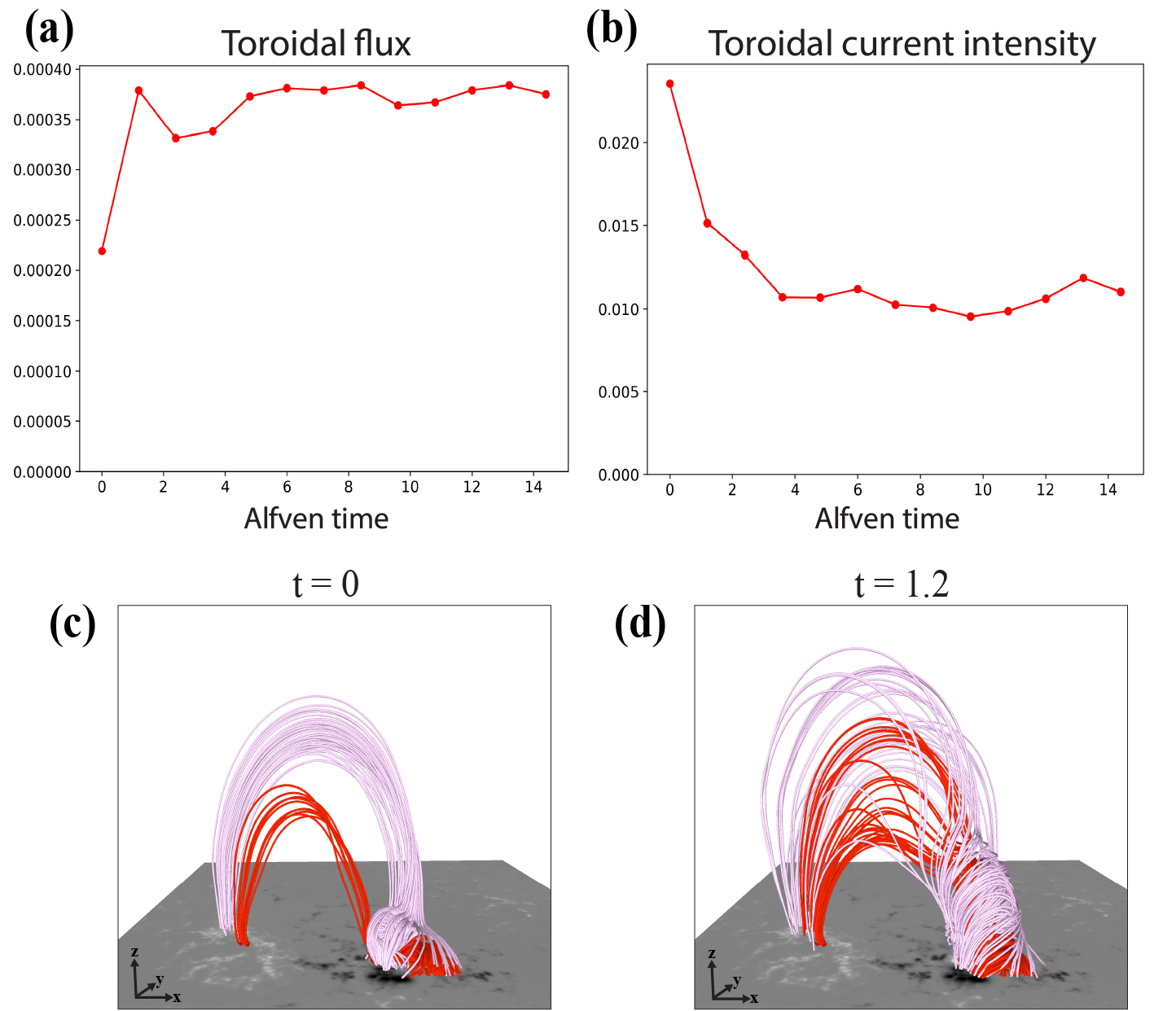} 
\caption{(a) and (b) Temporal evolution of the toroidal flux ($\int |B_z| dS \Big|_{T_w > 1.0}$) and toroidal current intensity ($\int |J_z| dS \Big|_{T_w > 1.0}$) of the MFR, respectively. The flux rope was identified based on a threshold twist
number of one full turn ($T_w > 1.0$). 
(c) and (d) 3D magnetic field line structure during Run 2d at $t=0$ and $t=1.2$, respectively, showing drastic changes in the magnetic topology.}
\label{fig:fig_11}
\end{figure}

Another important observational signature associated with CMEs is coronal dimming, characterized by transient regions of decreased emission in the EUV and soft X-ray (SXR) wavelengths. These dimmings are generally attributed to reduced plasma density caused by the expansion of the CME and the associated magnetic field restructuring (see Cheng \& Qiu \citeyear{cheng_nature_2016} and references therein). \citet{prasad_magnetohydrodynamic_2020} identified three primary coronal dimming regions in this active region (see Figure 2 in their paper). However, they did not determine the mechanism behind the formation of the dimming region indicated in Figure~\ref{fig:fig_9}(a) by a white arrow. Expanding on their study, we analyze this dimming region by comparing its location with the MFR footpoints in our study. Dissauer et al. (\citeyear{dissauer_projection_2016}) categorized this area as a potential `core dimming', which is typically interpreted as the footpoint of an erupting flux rope in most cases. Consequently, we propose that this dimming region corresponds to the westward footpoint of the expanding MFR as reproduced by our simulation. To verify, we overplot the MFR on the AIA 171 \AA\ image in Figure~\ref{fig:fig_9}(b), while Figure~\ref{fig:fig_9}(c) offers a side-view perspective that zooms in on the dimming region. The overlap between the MFR footpoint and the observed dimming region, as shown in Figures~\ref{fig:fig_9}(b) and ~\ref{fig:fig_9}(c), strengthens the validity of our results.



\section{Discussion} \label{sec:discussion}

Our simulation produced the large MFR through reconnection between the twisted sigmoidal fields, the adjacent outer fan-dome fields and the large coronal loops. The structure and evolution of this MFR, into which the fan–spine system becomes fully integrated, offer a coherent explanation for the observed circular flare ribbons and the associated coronal dimming region. Although the newly formed MFR exhibited an upward velocity, as seen in Figure~\ref{fig:fig_4}, a rapid eruption did not occur. To investigate this, we identify two contributing factors. The first is the limited size of the 3D computational box, which may have been insufficient to fully accommodate the large, expanding MFR. The second factor relates to the relaxation of the horizontal magnetic field components at the bottom boundary set in Run 2d. As noted earlier, when the fields are allowed to evolve, the twisted magnetic field lines may gradually relax, thereby weakening the hoop force that drives the MFR's upward motion.

\begin{figure}
\centering
\includegraphics[width=0.45\textwidth]{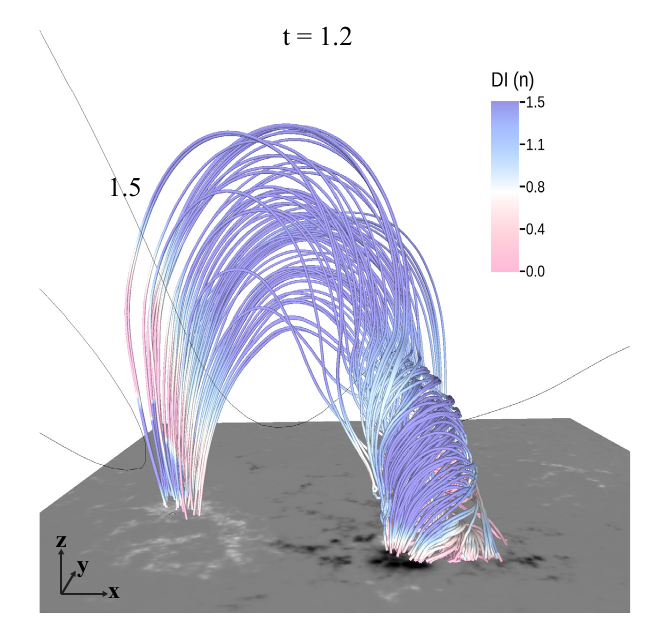} 
\caption{The decay index is plotted over the twisted magnetic field lines during Run 2d at $t = 1.2$. The black contour indicates the decay index, $n = 1.5$.}
\label{fig:fig_12}
\end{figure}

To assess the impact of the first factor, we calculated $\nabla \cdot \bm{V}$ throughout the region and plotted it on a vertical slice, as shown in Figure~\ref{fig:fig_10}(a). The color bar represents compressing regions (blue) and expanding regions (red). Panels (a)–(f) show that as the MFR rises, compressible waves propagate outward more rapidly and eventually collide with the side boundary around \textit{t} $\approx$ 10.8. These waves then reflect back, as seen between \textit{t} $\approx$ 12.0 and \textit{t} $\approx$ 13.2, eventually colliding with the MFR edge at \textit{t} $\approx$ 14.4. Based on this result, we suggest that this interaction likely suppresses the MFR’s ascent, contributing to its slower rise.

To address the second issue, we estimated the toroidal flux ($\int |B_z| dS \Big|_{T_w > 1.0}$) and the toroidal current intensity ($\int |J_z| dS \Big|_{T_w > 1.0}$) of the MFR as shown in Figures~\ref{fig:fig_11}(a) and ~\ref{fig:fig_11}(b), respectively. The MFR was identified based on a threshold twist number of one turn. Since the NLFFF lacks twisted field lines with $|T_w| \ge 1$, the MFR was defined as a bundle of newly twisted field lines formed via reconnection. Figure~\ref{fig:fig_11}(a) shows an initial rise in toroidal flux between times $t = 0$ and $t = 1.2$, which results from the redistribution of twist and toroidal current within the flux rope region between the two timesteps, as illustrated in Figures~\ref{fig:fig_11}(c) and ~\ref{fig:fig_11}(d). Notably, the fan-spine field lines become twisted and appear to be involved in the MFR by $t = 1.2$. As the system evolves, the toroidal flux eventually saturates, indicating that the total flux within the MFR becomes conserved. Meanwhile, Figure~\ref{fig:fig_11}(b) shows an initial decline in the current intensity, but at later stages, the slower expansion of the MFR appears to have led to a gradual change in the horizontal magnetic field, maintaining a nearly constant current density ($J_z$). 

\begin{figure*}
\centering
\includegraphics[width=0.97\textwidth]{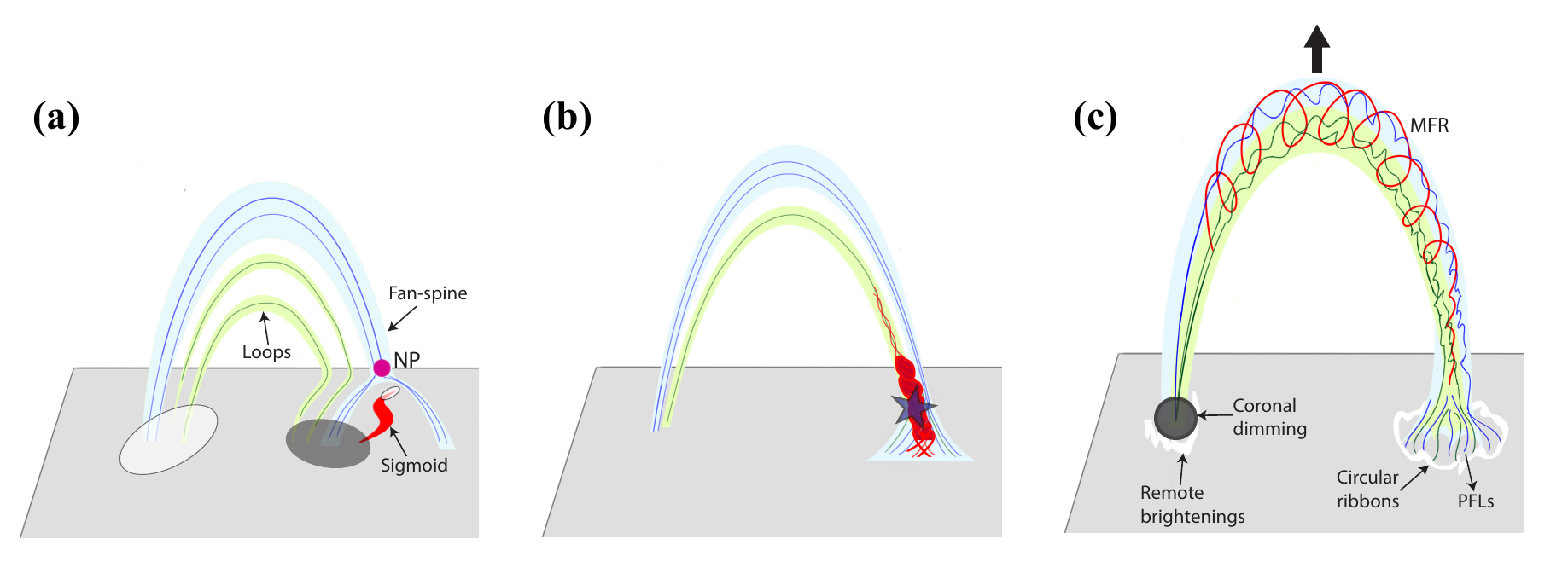} 
\caption{Schematic illustration of the flux rope evolution. (a) White and black ellipses represent positive and negative magnetic polarities, respectively. The S-shaped red line indicates the sigmoid; the thick blue lines outline the fan-spine structure with the NP marked in pink. The green lines correspond to neighboring loops that lie outside the fan dome. (b) Magnetic reconnection, marked by a star, occurs between the twisted red field lines and the green loops—part of the outer fan system. (c) Twisted field lines illustrate the propagation of twist within the expanding magnetic flux rope, formed via reconnection between the sigmoid, the fan-spine field lines and the neighboring loops. The westward footpoint of the flux rope is anchored at the circular flare ribbons along with post-flare loops (PFLs), while the eastward footpoint corresponds to remote brightenings and a dark-shaded region indicating coronal dimming. The black arrow denotes the direction of the MFR rise.}
\label{fig:fig_13}
\end{figure*}

To assess the stability of the MFR against torus instability (TI; \citealt{PhysRevLett.96.255002}), we compute the decay index, $n$, for Run 2d. The decay index quantifies how rapidly the strength of the external magnetic field (surrounding the MFR) decreases with height and is defined as:

\begin{equation}
\label{eq:DI}
n(x,y,z) = -\frac{d \textrm{ ln } |\bm{B_{\textrm{ext}}}(x,y,z)|}{d \textrm{ ln } z},
\end{equation}

where $z$ is the height above the photosphere, and $B_{ext}$ is the horizontal component of the external magnetic field. In this study, we assumed that the external field is the potential field \citep{2010ApJ...708..314A}. A value of $n \geq n_{cri} = 1.5 $ indicates susceptibility to torus instability, potentially enabling the MFR to erupt. 
Figure~\ref{fig:fig_12} shows the initial twisted magnetic structure colored by the decay index $n$, The field lines lie within regions where $n\geq1.5$, outlined by the black contour, suggesting that the flux rope has reached the critical condition for torus instability. While this estimate is approximate, it indicates that TI is unlikely to be the limiting factor for the lack of rapid eruption in our simulation.

These results suggest that the limited size of the simulation box influences the dynamical evolution of the MFR, particularly in Run 2d, leading to a slowing of its rise and likely inhibiting a full eruption—rather than effects such as hoop force weakening or an insufficient decay index. Thus, the current simulation domain appears insufficient to capture the long-term evolution of the MFR.



\section{Summary} \label{sec:summary}

In this paper, we investigated the flare dynamics during the X2.1 flare produced in AR11283 on September 06, 2011, using NLFFF extrapolation and data-constrained MHD modeling. We focused on understanding the complexities in the magnetic field topology and drew conclusions from our simulation results based on AIA observations. Figure~\ref{fig:fig_13} presents a simplified diagram of the MFR dynamics along with key observational characteristics. We summarize our main findings as follows:

\begin{enumerate}[itemsep=0pt]

\item The pre-flare magnetic configuration of AR11283 consists of a twisted sigmoidal structure above the PIL, surmounted by a fan-spine structure and neighboring large coronal loops located close to the outer fan-spine lines (Figure \ref{fig:fig_13}(a)). We employ anomalous resistivity to induce a magnetic field reconfiguration. Specifically, magnetic reconnection between the core sigmoidal field lines and adjacent fan-related loops caused the transition of the field lines from a static to a dynamic state (Figure \ref{fig:fig_13}(b)). These results indicate that reconnection within the strong current region plays a crucial role in triggering the early flare dynamics.

\item Magnetic reconnection facilitated a successive transfer of twist from low-lying, highly twisted sigmoidal field lines to other magnetic structures in the region, resulting in the formation and rise of a large MFR (Figure \ref{fig:fig_13}(c)). While our analysis qualitatively shows the redistribution of twist from the sigmoidal structure to the larger MFR, a quantitative assessment of the total end-to-end twist (e.g., using the method of \citet{2017ApJ...840...40G}) has not been performed in this study and is left for future work.

\item Previous studies have suggested that circular flare ribbons are likely associated with the fan–spine structure; however, the detailed dynamics of how the magnetic field is reconfigured during a flare have not been fully addressed. Our simulation reveals that the footpoints of the erupting flux rope align closely with the circular flare ribbon and a distant remote brightening region, offering a consistent explanation for their observed spatial separation.

\end{enumerate}

Our simulation successfully reproduced several phenomena observed in the X2.1 flare. However, the final MFR approached the full size of the computational domain, causing compressible waves to reflect off the side boundaries and interact with the MFR. Consequently, we were unable to track its long-term evolution or fully capture key physical processes such as its acceleration. To resolve this, a larger simulation box may be necessary to minimize boundary reflection effects and more accurately model the global eruption environment, particularly for inclined eruptions.



\begin{acknowledgments}

We are grateful to the referee for the valuable comments and suggestions that helped improve this paper. Data and images are courtesy of NASA/SDO and the HMI and AIA science teams. This study is supported by NSF grants AST-2204384 AGS-2145253 (CAREER), 2309939, 2401229, and 2408174, as well as NASA grants 80NSSC24M0174 (MIRO), 80NSSC23K0406 (HGI), 80NSSC21K1671 (HSR), 80NSSC24K1914, and 80NSSC21K0003 (LWS).
All numerical calculations in this paper were performed using the computing facilities of the High Performance Computing Center (HPCC) at the New Jersey Institute of Technology. The 3D visualizations were produced using VAPOR (\href{http://www.vapor.ucar.edu}{\texttt{www.vapor.ucar.edu}}), a product of the National Center for Atmospheric Research (\citealt{2019Atmos..10..488L}).
\end{acknowledgments}


\section*{ORCID iDs}
\noindent
Arpita Roddanavar: \href{https://orcid.org/0000-0003-1363-3096}{0000-0003-1363-3096} \\
Satoshi Inoue: \href{https://orcid.org/0000-0001-5121-5122}{0000-0001-5121-5122} \\
Keiji Hayashi: \href{https://orcid.org/0000-0001-9046-6688}{0000-0001-9046-6688} \\
Ju Jing: \href{https://orcid.org/0000-0002-8179-3625}{0000-0002-8179-3625} \\
Wenda Cao: \href{https://orcid.org/0000-0003-2427-6047}{0000-0003-2427-6047} \\
Haimin Wang: \href{https://orcid.org/0000-0002-5233-565X}{0000-0002-5233-565X}

\bibliographystyle{aasjournal}
\bibliography{Bibliography}{}

\end{document}